# Heterogeneous Employment Effects of Job Search Programmes: A Machine Learning Approach

Michael C. Knaus[†], Michael Lechner[‡], and Anthony Strittmatter[*]

*Date this version has been printed:* **26 February 2018**

**Abstract:** We systematically investigate the effect heterogeneity of job search programmes for unemployed workers. To investigate possibly heterogeneous employment effects, we combine non-experimental causal empirical models with Lasso-type estimators. The empirical analyses are based on rich administrative data from Swiss social security records. We find considerable heterogeneities only during the first six months after the start of training. Consistent with previous results in the literature, unemployed persons with fewer employment opportunities profit more from participating in these programmes. Finally, we show the potential of easy-to-implement programme participation rules for improving average employment effects of these active labour market programmes.

---

[†] Michael C. Knaus is also affiliated with IZA, Bonn.

[‡] Michael Lechner is also affiliated with CEPR, London, CESIfo, Munich, IAB, Nuremberg, and IZA, Bonn.

[*] Financial support from the Swiss National Science Foundation (SNSF) is gratefully acknowledged. The study is part of the project "Causal Analysis with Big Data" which has grant number SNSF 407540_166999 and is included in the Swiss National Research Programme "Big Data" (NRP 75). We presented a previous version of this paper at the Humboldt-University Berlin, Cal Poly in San Luis Obispo, University of Maastricht, Workshop on Unemployment and Labour Market Policies in Barcelona, the IZA Summer School in Labor Economics in Ammersee, Institute for Employment Research (IAB) in Nuremberg, and Computational Social Science Workshop in Konstanz. We thank participants, and in particular we thank Hugo Bodory, Bruno Crépon, Carlos Flores, Bernd Fitzenberger, Chris Hansen, Jeff Smith, and Martin Spindler, for helpful comments and suggestions. The usual disclaimer applies.

# 1 Introduction

In this study, we employ machine learning methods for a systematic investigation of effect heterogeneity of job search programmes ('JSPs' from now on) in Switzerland. Programme evaluation studies widely acknowledge the possibility of effect heterogeneity for different groups. Stratifying the data in mutually exclusive groups or including interactions in a regression framework are two baseline approaches to investigate effect heterogeneity (see, e.g., Athey and Imbens, 2017a, for a review). However, these approaches may overlook important heterogeneities because they usually do not include a *systematic* search based on clear, spelled-out statistical rules. Furthermore, for large-scale investigations of effect heterogeneity, standard p-values of classical (single) hypothesis tests are no longer valid because of the multiple-hypothesis testing problem (see, e.g., Lan et al., 2016, List, Shaikh, and Xu, 2016). For example, for fifty single hypotheses tests, the probability that at least one test falsely rejects the null hypotheses at the 5% significance level could be up to 92%.[1] This could lead to so-called *ex post selection* and the reporting of spurious heterogeneity that, in fact, resulted from so-called false positives.

The disadvantages of *ex post selection* of significant effects have been widely recognized in the programme-evaluation literature. For example, in randomized experiments researchers may be required to define their analysis plan for heterogeneity prior to the experiment to avoid only reporting (and searching for) significant effects (e.g., Casey, Glennerster, and Miguel, 2012, Olken, 2015). However, these pre-analysis plans are inflexible and usually not demanded (by funding bodies or editors of economic journals) in the common case of observational studies. An alternative approach that partly alleviates the *ex post selection* problem is to report effect heterogeneity for all possible groups. For large-scale investigations, an approach that takes account of all possible differences might lead to very small groups and thus imprecise estimates. Further, the large number of different results makes it difficult to report the results in an intuitive way.

---

[1] Assuming independent test statistics as an extreme case $(1 - 0.95^{50} = 0.92)$.

A developing part of the literature proposes to use machine learning algorithms (adapted for causal analysis) to systematically search for groups with heterogeneous effects (see, e.g., the review of Athey and Imbens, 2017b). Machine learning approaches are potentially attractive because they could provide a principled approach to heterogeneity detection, which make it less likely to leave out important heterogeneities and can reduce concerns about the multiple testing problem. In addition, they enable flexible modelling and remain computationally feasible, even when the covariate space becomes high-dimensional and possibly exceeds the sample size.

This is one of the first large-scale economic application of causal machine learning methods investigating effect heterogeneity for observational studies. In contrast to the few existing studies, which mainly exploit randomised control trials (e.g., Bertrand et al., 2017, Denisova-Schmidt et al., 2017), we pursue the more common but also more demanding selection-on-observables strategy (see, e.g., Imbens and Wooldridge, 2009, for standard econometric approaches and their properties). For this reason, we exemplify the application of causal machine learning methods at the example of a well-studied training programme for which the evaluation literature tackled the selection issues plausibly (see e.g., e.g. Card, Kluve, and Weber, 2015, for an overview and meta analysis of evaluation studies of active labour market programmes). This enables us to focus on the econometric innovations of the causal machine learning approach.

In particular, we contribute to the causal machine learning literature in at least two ways. First, we systematically investigate effect heterogeneity of JSPs and report them in an interpretable way. We base the search algorithm for heterogeneity on many attributes of the unemployed persons as well as their caseworkers. For example, we consider the employment and welfare history of unemployed persons, socio-demographic characteristics, caseworkers' subjective employability ratings of their clients, and measures for the cooperativeness of caseworkers. The latter could uncover effect heterogeneity by different monitoring intensities, which we consider an important mechanism of JSPs (Behncke, Frölich, and Lechner, 2010a). Overall, we consider 1,268 different variables, including interactions and polynomials. Second, based on the detected heterogeneities, we document the potential of different assignment rules to improve JSPs' effects and cost-benefit efficiency.

Importantly, we investigate the consistency of our findings across a large variety of different machine learning algorithms. The causal machine learning literature is still lacking large-scale sensitivity checks with regard to methodological choices in credible applications; particularly in observational studies. Obviously, the robustness of the results to possible misspecifications of the empirical model is essential for drawing coherent policy conclusions.

For the investigation of effect heterogeneity, we combine Inverse Probability Weighting (IPW) with the so-called Modified Covariate Method (MCM) (Tian et al., 2014, Chen et al., 2017). We use exceptionally rich linked unemployed-caseworker data obtained from Swiss social security records. The selection of relevant heterogeneity is carried out with Tibshirani's (1996) Least Absolute Shrinkage and Selection Operator (LASSO). For the quantification of the effects and their inference, we follow the sample splitting approach (see recent discussion in Rinaldo et al., 2016). We use half of the sample to select variables that are relevant to predict the size of the heterogeneous treatment effect, i.e. that are responsible for deviations from the average effects. We use the other half of the sample for inference on the (possibly low-dimensional) selected variables and the heterogeneous effects.

Our results suggest substantial effect heterogeneity of Swiss JSPs during the first six months after the start of participation. During this so-called 'lock-in' period, we observe negative effects for most participants. However, the size of the heterogeneity is strongly related to the characteristics of the unemployed. Consistent with the previous literature, participants with disadvantaged labour market characteristics benefit more from JSPs (e.g., Card, Kluve, and Weber, 2015). A major reason is that they face generally lower lock-in effects and, thus, these indirect programme costs are lower. Additionally, this study appears to be the first to uncover substantial effect heterogeneity by residence status. We show that JSPs are more effective for foreigners, who have less access to informal job search networks compared to locals. For caseworker characteristics, however, there is only little heterogeneity. There is also no substantial effect heterogeneity beyond six months after the start of training. Finally, the paper presents easy-to-implement assignment rules which would improve the current assignment mechanism in a (almost) cost-neutral way. An extensive sensitivity analysis shows that the main conclusions remain robust across a variety of different estimation methods.

In the next section, we provide information about the institutional background of the Swiss ALMP. In Section 3, we document the sample selection and show basic descriptive statistics. In Section 4, we discuss the econometric approach for a principled investigation of effect heterogeneity. In Section 5, we report the empirical findings and robustness checks. Section 6 explains our conclusions. Additional descriptive statistics, detailed information on the selection procedures, results for additional outcome variables, as well as extensive sensitivity analyses are reported in Online Appendices A-F.

## 2 Background

### 2.1 Swiss institutions

Switzerland is a federal country with 26 cantons and three major language regions (French, German, and Italian). It is a relatively wealthy country with approximately 78,000 CHF (approx. 77,000 US-Dollar) GDP per capita and a low unemployment rate of 3 to 4% (SECO, 2017, Federal Statistical Office, 2017). Unemployed persons have to register at the regional employment agency closest to their home.[2] The employment agency pays income maintenance. Benefits amount to 70 to 80% of the former salary depending on age, children, and past salary (see Behncke, Frölich, and Lechner, 2010b). The maximum benefit entitlement period is 24 months.

The yearly expenditures for Swiss ALMPs exceed 500 million CHF (Morlok et al., 2014). Unemployed persons can participate in a variety of different ALMPs. Gerfin and Lechner (2002) classify these ALMPs as (a) training courses, (b) employment programmes, and (c) temporary employment schemes. Training courses include job search, personality, language, computer, and vocational programmes. We focus on JSPs in this study, which is the most common ALMP in Switzerland (more than 50% of the assigned ALMPs are JSPs, see Huber, Lechner, and Mellace, 2017). JSPs provide training in effective job search and application strategies (e.g., training in résumé writing). Furthermore, actual applications are screened and monitored. JSPs are relatively short, with an average duration of about three weeks. Training takes place in classrooms. The employment agency covers the costs of training and travel. Participants are obliged to continue to search for jobs during the course.

In Switzerland, regional employment agencies have a large degree of autonomy, which is partly related to the country's federal organisation. Caseworkers make the decision to assign unemployed persons to a training course based on information about the unemployed person (e.g. employment history, subjective employability rating, etc.). Additionally, employment agency policies and federal eligibility rules are relevant for the assignment decision. The federal eligibility rules are rather vague. They imply, for example, that the training has to be necessary and adequate to improve the individual's employment chances. Caseworkers can essentially force the unemployed into such courses by threatening to impose sanctions. Unemployed persons have the option to apply to participate in such courses, but the final decision is always made by the caseworkers.

---

[2] At the beginning of the unemployment spell, newly registered unemployed persons are often sent to a one-day workshop providing information about the unemployment law, obligations and rights, job search requirements, etc.

## 2.2   Related literature on job search programmes (JSPs)

An assignment to a JSP may affect the matching process and quality alignment between the participant and his or her potential new job (see, e.g., Blasco and Rosholm, 2011, Cottier et al., 2017). Push effects could occur if participants accept jobs with low matching quality because of actual or perceived sanctions or perceived future ALMP assignments. Push effects decrease the duration of unemployment, but may reduce employment stability. On the other hand, JSP participation could improve the visibility of suitable job vacancies and the efficiency of the application process, which may improve employment stability. Furthermore, many studies are concerned with the crowding-out of non-participants (see, e.g., Blundell et al., 2004, Crépon et al., 2013, Gautier et al., 2017).

Empirical evidence about the effectiveness of JSPs is mixed. The review studies of Card, Kluve, and Weber (2010, 2015) as well as Crépon and van den Berg (2016) document a weak tendency towards positive effects of JSPs, especially in the short-term.[3] However, for Swiss JSPs, the literature finds negative employment effects, which taper off one year after the start of participation (see Gerfin and Lechner, 2002, Lalive, van Ours, and Zweimüller, 2008). One reason for the ambiguous effectiveness of JSPs might be the different relative intensities of job search training and monitoring. Van den Berg and van der Klaauw (2006) are concerned that intensive monitoring reduces informal job search, which might be a more efficient strategy than formal job search for some unemployed persons. They suggest formal job search is more effective for individuals with fewer labour market opportunities. Consistent with their arguments, Card, Kluve, and Weber (2015) document that JSPs are relatively more effective for disadvantaged participants. Vikström, Rosholm, and Svarer (2013) find slightly more positive effects of JSPs for women and younger participants. Dolton and O'Neill (2002) report negative employment effects of JSPs for men and insignificant effects for women five years after the programme begins. Surprisingly, the programme evaluation literature is lacking large-scale evidence about the effect heterogeneity of JSPs.

---

[3] Meyer (1995) reports negative effects on unemployment benefit payments and positive earnings effects of JSPs in the US. Graversen and van Ours (2008) and Rosholm (2008) report positive effects of JSPs on the unemployment exit rate in Denmark. Wunsch and Lechner (2008) find that JSPs have negative effects during the first two years after a programme begins, which fade out afterwards in Germany. They also show that training sequences are responsible for long-lasting negative lock-in effects.

# 3 Data

## 3.1 General

The data we use includes all individuals who are registered as unemployed at a Swiss regional employment agency in the year 2003. The data contains rich information from different unemployment insurance databases (AVAM/ASAL) and social security records (AHV). This is the standard data used for many Swiss ALMP evaluations (e.g. Gerfin and Lechner, 2002, Lalive, van Ours, and Zweimüller, 2008, Lechner and Smith, 2007). We observe (among others) residence status, qualification, education, language skills, employment history, profession, job position, industry of last job, and desired occupation and industry. The administrative data is linked with regional labour market characteristics, such as the population size of municipalities and the cantonal unemployment rate. The availability of extensive caseworker information and their subjective assessment of the employability of their clients is what distinguish our data. Swiss caseworkers employed in the period of 2003 to 2004 were surveyed through a written questionnaire in December 2004 (see Behncke, Frölich, and Lechner, 2010a, 2010b). The questionnaire asked about the caseworker's aims and strategies and information about the regional employment agency.

## 3.2 Sample definition

In total, 238,902 persons registered as being unemployed in 2003. We only consider the first unemployment registration per individual in 2003. Each registered unemployed person is assigned to a caseworker. In most cases, the same caseworker is responsible for the entire unemployment duration of his or her client. If this is not the case, we focus on the first caseworker to avoid concerns about (rare) endogenous caseworker changes (see Behncke, Frölich, and Lechner, 2010b). We only consider unemployed persons aged between 24 and 55 years who receive unemployment insurance benefits. We omitted unemployed persons who apply for disability insurance benefits, persons whose responsible caseworker is not clearly defined, or persons whose caseworker did not answer the questionnaire (the response rate is 84%). We omitted unemployed foreigners with a residence permit that is valid for less than a year. Finally, we omitted unemployed persons from five regional employment agencies that are not comparable to the other regional employment agencies. This sample is identical to the data used in Huber, Lechner, and Mellace (2017). It contains 100,120 unemployed persons.

One concern regarding the treatment definition is the timing with respect to the elapsed unemployment duration prior to participation. Caseworkers may assign unemployed persons to job training programmes at essentially anytime during their unemployment spell. Such dynamic or sequential programme assignment has received considerable attention in the evaluation literature (see the discussions in Abbring and van den Berg, 2003, 2004, Fredriksson and Johansson, 2008, Heckman and Navarro, 2007, Lechner, 2009, Robins, 1986, Sianesi, 2004, among others). We consider a classical static evaluation model and define treatment as the first participation in a JSP during the first six months of unemployment (83% of JSP are assigned within the first six months of unemployment). We exclude individuals who participate in other ALMPs within the first six months of unemployment from the sample, such that our control group represents non-participants of all programmes (8,787 other ALMP participants are dropped). Potentially, this approach could lead to a higher share of individuals with better labour market characteristics among the control group than among the training participants, because individuals in the control group may have possibly found another job prior to their potential treatment times. This would negatively bias the results. To overcome this concern, we assign conditional random (pseudo) participation starts to each individual in the control group. Thereby, we recover the distribution of the elapsed unemployment duration at the time of training participation from the treatment group (similar to, e.g., Lechner, 1999, Lechner and Smith, 2007). To ensure comparability of the treatment definitions of the participants and non-participants, we only consider individuals who are unemployed at their (pseudo) treatment dates. This makes the groups of participants and non-participants comparable with respect to the duration of unemployment and ensures that the treated and control groups are eligible for programme participation at their respective assigned start dates.

The final sample contains 85,198 unemployed persons (Table A.1 in Online Appendix A provides the details of the sample selections steps). From this sample, 12,998 unemployed persons participate in a JSP and 72,200 are members of the control group. These 85,198 unemployed persons are assigned to 1,282 different caseworkers.

### 3.3 Descriptive statistics

Table 1 reports the means and standard deviations by JSP participation for some selected variables. During the first 6 months after training begins, JSP participants are fewer months employed than non-participants. The standardised difference is above 20.[4] During the first 12 and 31 months after training begins, JSP participants also have a shorter employment duration

---

[4] The standardised difference of variable $X$ between samples $A$ and $B$ is defined as

$$SD = \frac{|\bar{X}_A - \bar{X}_B|}{\sqrt{1/2\left(Var(X_A) + Var(X_B)\right)}} \cdot 100,$$

where $\bar{X}_A$ denotes the mean of sample $A$ and $\bar{X}_B$ denotes the mean of sample $B$. Rosenbaum and Rubin (1983) consider a standardised difference of more than 20 as being 'large'.

than non-participants, but the standardised differences decline. During months 25 to 31 after training begins, the difference in the employment duration is minor.

*Table 1: Descriptive statistics of some important variables by JSP participation status.*

|  | Participants | | Non-Participants | | Std. Diff. |
|---|---|---|---|---|---|
|  | Mean | S.D. | Mean | S.D. |  |
|  | (1) | (2) | (3) | (4) | (5) |
| Outcome: Months employed since programme start | | | | | |
| During first 6 months | 1.21 | 1.93 | 1.94 | 2.44 | 23.29 |
| During first 12 months | 3.68 | 4.27 | 4.53 | 4.80 | 13.12 |
| During first 31 months | 15.30 | 12.49 | 15.59 | 12.85 | 1.60 |
| During months 25 - 31 | 3.48 | 2.88 | 3.33 | 2.86 | 3.72 |
| Characteristics of unemployed persons | | | | | |
| Female | 0.45 | - | 0.44 | - | 0.58 |
| Age (in 10 years) | 3.73 | 0.88 | 3.66 | 0.86 | 5.59 |
| Unskilled | 0.22 | - | 0.23 | - | 1.80 |
| Some qualification degree | 0.60 | - | 0.56 | - | 5.19 |
| Employability rating low | 0.12 | - | 0.14 | - | 3.97 |
| Employability rating medium | 0.77 | - | 0.74 | - | 5.79 |
| Employability rating high | 0.11 | - | 0.12 | - | 3.62 |
| # of unemp. spells in last 2 years | 0.41 | 0.98 | 0.64 | 1.27 | 13.85 |
| Fraction of months emp. in last 2 years | 0.83 | 0.22 | 0.79 | 0.25 | 12.57 |
| Past income (in 10,000 CHF) | 4.58 | 2.02 | 4.16 | 2.05 | 14.50 |
| Caseworker characteristics | | | | | |
| Female | 0.45 | - | 0.41 | - | 6.94 |
| Age (in years) | 44.0 | 11.6 | 44.4 | 1.16 | 7.7 |
| Tenure (in years) | 5.54 | 3.23 | 5.86 | 3.31 | 6.84 |
| Own unemp. experience | 0.63 | - | 0.63 | - | 0.54 |
| Vocational training degree | 0.26 | - | 0.23 | - | 5.63 |
| Local labour market characteristics | | | | | |
| German speaking REA | 0.89 | - | 0.67 | - | 39.68 |
| French speaking REA | 0.08 | - | 0.25 | - | 33.30 |
| Italian speaking REA | 0.03 | - | 0.08 | - | 16.81 |
| Cantonal unemployment rate (in %) | 3.64 | 0.77 | 3.75 | 0.86 | 9.23 |
| Cantonal GDP per capita (in 10,000 CHF) | 5.13 | 0.92 | 4.92 | 0.93 | 15.75 |
| # of caseworkers | 989 | | 1,282 | | |
| # of observations | 12,998 | | 72,200 | | |

Note: We report unconditional means for all variables, standard deviations (S.D.) for all non-binary variables, and standardised differences between participants and non-participants. The descriptive statistics of all confounding variables used in this study are shown in Table B.1 of Online Appendix B. REA is the abbreviation for regional employment agency.

Furthermore, Table 1 documents descriptive statistics of the characteristics of the unemployed person, the characteristics of his or her caseworker, and local labour market conditions. We report descriptive statistics for additional control variables in Table B.1 of Online Appendix B. JSP participants have spent more months employed and received a higher income than non-participants in the last two years prior to the programme's start. We document minimal difference between the caseworkers of participants and non-participants.[5] JSP participants are more often registered at German-speaking regional employment agencies and live in cantons with better economic conditions (in terms of local GDP and unemployment rate) than non-participants.

# 4 Econometric approach

## 4.1 Parameters of interest

We describe the parameters of interest using Rubin's (1974) potential outcome framework. Following the conventional notation, we indicate random variables by capital letters and the realizations of these random variables by lowercase letters. The binary treatment dummy $D_i$ indicates JSP participation. Let $Y_i^1$ denote the potential outcome (e.g., employment) when individual $i$ ($i = 1, \ldots, N$) participates in a JSP ($D_i = 1$). Conversely, $Y_i^0$ denotes the potential outcome when individual $i$ is not participating in a JSP ($D_i = 0$). Obviously, each individual can either participate in a JSP or not, but both participation states cannot occur simultaneously. This implies only one potential outcome is observable. The observed outcome equals

$$Y_i = Y_i^1 D_i + Y_i^0 (1 - D_i).$$

The causal effect of $D$ on $Y$ for individual $i$ is

$$\gamma_i = Y_i^1 - Y_i^0.$$

---

[5] In Table B.1 of Online Appendix B we also show caseworker characteristics interacted with the language of the regional employment agency. For some interacted variables, we find strong differences between participants and non-participants.

However, we cannot identify the parameter $\gamma_i$ without assumptions that are implausible in many applications (e.g., effect homogeneity). Nevertheless, group averages of $\gamma_i$ may be identifiable under plausible assumptions. For example, the identification of the average treatment effect (ATE), $\gamma = E[\gamma_i]$, the average treatment effect on the treated (ATET), $\theta = E[\gamma_i|D_i = 1]$, and the average treatment effect on the non-treated (ATENT), $\rho = E[\gamma_i|D_i = 0]$, are standard econometric problems (see, e.g., Imbens and Wooldridge, 2009). Furthermore, conditional average treatment effects (CATEs) can potentially uncover effect heterogeneity based on exogenous pre-treatment variables $Z_i$ chosen by the researcher based on the policy interest,

$$\gamma(z) = E[\gamma_i|Z_i = z] = E[Y_i^1 - Y_i^0|Z_i = z].$$

Knowledge about CATEs could help, e.g., to improve the assignment mechanism to JSPs.[6]

## 4.2 Econometric background and intuition

Machine learning methods are powerful tools for out-of-sample predictions of observable variables. However, the fundamental problem of causal analyses is the inability to observe individual causal effects because at least one potential outcome is unobservable. Recently, several methods have been proposed that apply machine learning methods in ways that overcome this fundamental problem (see, e.g., the reviews by Belloni, Chernozhukov, and Hansen, 2014, Horowitz, 2015, and Varian, 2014).

Concerning effect heterogeneity, Imai and Ratkovic (2013) suggest a LASSO-type algorithm while Athey and Imbens (2016) propose a regression tree method. Foster, Taylor, and Ruberg (2011) apply random forest algorithms to estimate effect heterogeneity. These algorithms are flexible and are effective at capturing multi-dimensional and non-linear interactions among covariates. Imai and Strauss (2011), Green and Kern (2012), and Taddy et al. (2015) propose alternative Bayesian machine learning methods to estimate effect heterogeneity. Grimmer, Messing, and Westwood (2017) do not attempt to use the best method available. Instead, they suggest combining many different machine-learning tools to estimate the conditional treatment responses. Athey and Wager (2017), Qian and Murphy (2011), Xu et al. (2015), and Zhao et al. (2012) focus on the estimation of individualized treatment rules, which primarily focus on decision rules instead of effect heterogeneity.[7]

---

[6] Additional parameters are CATEs for JSP participants $\theta(z) = E[\gamma_i|D_i = 1, Z_i = z]$ and CATEs for non-participants $\rho(z) = E[\gamma_i|D_i = 0, Z_i = z]$. The parameters $\gamma(z)$, $\theta(z)$, and $\rho(z)$ can differ from each other when $Z_i$ differs from $X_i$ (which is the case in our application). However, we are interested in the heterogeneities for a random unemployed person with specific characteristics because this mirrors the decision problem of the caseworker. Thus, we focus on $\gamma(z)$.

All of these studies consider heterogeneity in randomized experiments. In many fields of economics, randomized experiments are expensive and minimally socially acceptable. Therefore, we consider a selection-on-observables identification strategy (e.g. Imbens, Wooldridge, 2009). A promising approach to estimate group specific causal effects in non-experimental contexts is the Modified Covariate Method (MCM).[8]

To gain some intuition about the MCM, we consider the stylised case where participation in a programme is randomly assigned to 50% of the unemployed persons. Accordingly, in this introductory example there is no need to adjust for selection into training participation. Throughout the analyses the first element in $Z_i$ is a constant term ($Z_{i0} = 1$) and the remaining elements of $Z_i$ contain additional $p \geq 1$ pre-treatment variables that are potentially related to the effect heterogeneity in which the researcher is interested. A standard approach to estimate effect heterogeneity is to use the interaction model,

$$Y_i = Z_i \beta_s + D_i Z_i \delta + u_i. \qquad (1)$$

---

[7] Closely related is the study of Ciarleglio et al. (2015), who propose a method to select the optimal treatment conditional on observed individual characteristics. Zhao et al. (2015) investigate the optimal dynamic order of sequential treatments.

[8] Furthermore, Zhang et al. (2012) develop alternative non-experimental approaches for a principled effect heterogeneity search, which is an adaptation of the Modified Outcome Method (MOM) (Signorovitch, 2007). We describe the MOM in Online Appendix F.1. For one of the robustness checks, we replicate our results using the MOM. Furthermore, the tree and forest methods of Athey and Imbens (2016) and Wager and Athey (2017) are applicable in non-experimental settings. All robustness checks are provided in Section 5.7 and Online Appendix F. The main findings are not altered.

The first term on the right side of equation (1) provides a linear approximation of the conditional expectation of the potential outcome under non-participation, $E[Y_i^0|Z_i = z] = z\beta_s$. We call this the main effects in the following. The second term on the right-hand side of equation (1) provides a linear approximation of the CATE:

$$\gamma(z) = z\delta = E[Y_i^1 - Y_i^0|Z_i = z].$$

Vansteelandt et al. (2008) point at possible sensitivities of the empirical model in equation (1) when the main effects are mis-specified. Tian et al. (2014) propose to transform the treatment dummy $T_i = 2D_i - 1$ and rearrange the interaction model in equation (1) to:

$$Y_i = Z_i\beta_t + \frac{T_iZ_i}{2}\delta + v_i. \tag{2}$$

The treatment indicator shifts from $D_i \in \{0,1\}$ to $T_i/2 \in \{-0.5, 0.5\}$. The modification does not alter the coefficient vector $\delta$. However, this transformation alters the main effects. In equation (2), $E[Y_i|Z_i = z] = z\beta_t$ is the linear approximation of the conditional expectation of the observed outcome. Notice that $Cov(Z_{ij}, T_iZ_{ik}) = Cov(Z_{ij}, Z_{ik})E[T_i] = 0$ for $j, k \in \{1, ..., p\}$. The first equality holds under random assignment of training participation and the second equality holds because $E[T_i] = 0$.[9] Accordingly, the right hand terms of equation (2) are independent of each other and we can estimate the coefficients $\beta_t$ and $\delta$ in two separate regressions. For example, we can estimate CATEs with the model

$$Y_i = \frac{T_iZ_i}{2}\delta + \varepsilon_i,$$

which is the baseline model of the MCM. The MCM is suitable when only the interaction effects and not the main effects are of interest. Parsimony and robustness to misspecification of the main effects are two advantages of the MCM compared to the specification in equation (1). We can adopt the basic idea of the MCM to non-experimental identification strategies (see Chen et al., 2017). Furthermore, we can combine the MCM with different machine learning methods to select the variables for heterogeneity. Procedure 1 summarises our (main) estimation algorithm of the adapted MCM approach, which we describe in detail below.

---

[9] In contrast, $Cov(Z_{ij}, D_iZ_{ik}) = Cov(Z_{ij}, Z_{ik})E[D_i] = Cov(Z_{ij}, Z_{ik})/2$, which can be different from zero.

## 4.3 Identification

In addition to the pre-treatment variables included in the vector $Z_i$ (which are potentially related to effect heterogeneity), we consider the possibility of confounding variables, which are included in the vector $X_i$. Confounders are pre-treatment variables that jointly affect the probability to participate in a JSP and the employment outcome. The vector $Z_i$ may be larger, smaller, partially, or fully overlapping with $X_i$ depending on the question under investigation.

**Assumption 1** (Conditional independence): $Y_i^1, Y_i^0 \coprod D_i | X_i = x, Z_i = z$ for all values of $x$ and $z$ in the support of $X$ and $Z$.

**Assumption 2** (Common support): $0 < P(D_i = 1 | X_i = x, Z_i = z) = p(X_i, Z_i) < 1$ for all values of $x$ and $z$ in the support (of interest) of $X$ and $Z$.

**Assumption 3** (Exogeneity of controls): $X_i^1 = X_i^0$ and $Z_i^1 = Z_i^0$.

**Assumption 4** (Stable Unit Treatment Value Assumption, SUTVA): $Y_i = Y_i^1 D_i + Y_i^0 (1 - D_i)$.

Assumption 1 states that the potential outcomes are independent of programme participation conditional on the confounding pre-treatment variables. The plausibility of this assumption is justified by the availability of a detailed set of confounding variables containing characteristics of the unemployed and the caseworkers. The studies of Biewen et al. (2014) and Lechner and Wunsch (2013) discuss the selection of confounders in ALMP evaluations based on rich administrative data. Within the employment agency, caseworkers have high autonomy to decide about assignment of JSPs. Our data contain the same objective measures about labour market history, education and socio-demographics of the unemployed, as well as local labour market characteristics that are observable to the caseworkers when choosing who participates in JSPs. We observe caseworkers' subjective ratings of the employability of their clients. Furthermore, we observe detailed information about the caseworkers' characteristics and counselling styles. These are potential confounders, because caseworker characteristics might affect the probability of JSP participation and labour market outcomes simultaneously.

According to Assumption 2, the conditional probability to participate in a JSP is bounded away from zero and one. The common support assumption has to hold when conditioning jointly on $X$ and $Z$. We enforce common support by trimming observations below the 0.5 quantile of participants and above the 99.5 quantile of non-participants.[10] This procedure shows good finite sample performance in the study of Lechner and Strittmatter (2017). Assumption 3 requires exogeneity of confounding and heterogeneity variables. To account for this assumption, we only use control variables that are determined prior to the start of JSP participation. Assumption 4 excludes spillover effects between participants and non-participants.

**Theorem 1** (Identification): Under Assumptions 1-4 (and regularity conditions ensuring the existence of appropriate moments) the following equality holds:

$$\gamma(z) = E_{X|Z=z}[E(Y_i| D_i = 1, X_i = x, Z_i = z)|Z_i = z]$$
$$- E_{X|Z=z}[E(Y_i| D_i = 0, X_i = x, Z_i = z)|Z_i = z].$$

Thus $\gamma(z)$ are identified from observable data on $\{Y_i, D_i, Z_i, X_i\}_{i=1}^N$. For completeness, the proof of Theorem 1 is in Online Appendix C (see also, e.g., Rosenbaum and Rubin, 1983).

## 4.4 Search for effect heterogeneity

Chen et al. (2017) outline how we can combine MCM with Inverse Probability Weighting (IPW), a standard approach to balance covariates in observational studies (see, e.g., Hirano,

---
[10] In total, we trim 6,767 observations (579 participants, 6,188 non-participants).

Imbens, and Ridder, 2003, Horvitz and Thompson, 1952). We can estimate the parameter vector $\delta$ using Weighted Ordinary Least Squares (WOLS), i.e. by minimising the objective function

$$\underset{\hat{\delta}}{\mathrm{argmin}}\left[\sum_{i=1}^{N}\hat{w}(D_i,X_i,Z_i)T_i\left(Y_i-\frac{T_iZ_i}{2}\delta\right)^2\right], \quad (3)$$

with the IPW weights

$$\hat{w}(D_i,X_i,Z_i) = \frac{\frac{D_i-\hat{p}(X_i,Z_i)}{\hat{p}(X_i,Z_i)(1-\hat{p}(X_i,Z_i))}}{D_i\sum_{i=1}^{N}\frac{D_i}{\hat{p}(X_i,Z_i)}+(1-D_i)\sum_{i=1}^{N}\frac{1-D_i}{1-\hat{p}(X_i,Z_i)}},$$

which we calculate using the estimated propensity score $\hat{p}(X_i,Z_i)$. In our baseline model, we adapt the propensity score specification of Huber, Lechner, and Mellace (2017), which we report in Table D.1 of Online Appendix D. The denominator of the IPW weights is a small sample adjustment to ensure that the weights for treated and controls sum to one (see, e.g., Busso, DiNardo, and McCrary, 2014). In equation (3), we multiply the IPW weights by $T_i$, such that the weights are positive.

The variables included in $Z_i$, which are potentially related to effect heterogeneity, consist of individual and caseworker characteristics, their second order interactions, up to fourth order polynomials, and logarithms of non-binary variables. Additionally, we consider dummy variables for the 103 employment agencies as well as 29 category dummies for previous industry and 29 category dummies describing the previous job. In total, this leads to 1,268 heterogeneity variables that we consider in the analyses.[11]

In our main specifications, we employ LASSO estimators. The weighted LASSO estimator of the MCM minimizes the objective function,

---

[11] We exclude binary variables where less than 1% of (non-) participants show values of 0 or 1. Furthermore, we keep only one variable of variable combinations that show correlations of larger magnitude than ±0.99 to speed up computation.

$$\underset{\hat{\delta}}{\mathrm{argmin}} \left[ \sum_{i=1}^{N} \widehat{w}(D_i, X_i, Z_i) T_i \left( Y_i - \frac{T_i Z_i}{2} \hat{\delta} \right)^2 \right] + \lambda \sum_{j=1}^{p} |\hat{\delta}_j|, \qquad (4)$$

where we add a penalty term for the sum of the absolute values of the coefficients of the $p$ variables appearing in $Z$. Importantly, we do not penalize the constant $\hat{\delta}_0$. The penalising parameter $\lambda$ specifies the amount of penalisation. If $\lambda = 0$, then equation (4) is equivalent to the WOLS model in equation (3). However, when $\lambda > 0$ some coefficients are shrunken towards zero. For sufficiently large values of $\lambda$, some (or all) coefficients are exactly zero. Therefore, the LASSO serves as a model selector, omitting variables with little predictive power from the model.[12] A challenge is the optimization of the penalty term, such that only the relevant predictors of the effect heterogeneity remain in the model. Too low penalties lead to overfitting, too high penalties lead to models that miss important variables (i.e., we have a bias-variance trade-off).

We apply 10-fold cross-validation to find the penalty term $\lambda$ with the best out-of-sample performance in terms of mean-squared-error (MSE) (e.g., Bühlmann and van de Geer, 2011).[13] The LASSO coefficients are biased when $\lambda > 0$ (regularisation bias, see, e.g., Zou, 2006). For this reason, we use the so-called Post-LASSO estimates to calculate the MSE. We obtain the Post-LASSO coefficients from a WOLS model, which includes all variables with non-zero coefficients in the respective LASSO model (see, e.g. Belloni and Chernozhukov 2013). We choose the LASSO model with the penalty parameter $\lambda$ that minimises the Post-LASSO MSE.[14]

There is no need to specify the main effects in the MCM approach. Nevertheless, Tian et al. (2014) and Chen et al. (2017) show that accounting for the main effects can improve the finite sample performance of the MCM because they can absorb variation in the outcome, which is unrelated to the effect heterogeneity. In Online Appendix F.2, we document two ways to implement an efficiency augmenting procedure.

---

[12] The larger the values of $\lambda$ the fewer variables remain in the model. By gradually increasing the penalty term one can obtain a path from a full model to a model that only contains the parameter $\hat{\delta}_0$.

[13] Chetverikov, Liao, and Chernozhukov (2017) discuss the properties of K-fold cross-validation in the context of LASSO. They derive bounds for the prediction errors of cross-validated LASSO estimators.

[14] In robustness checks, we base the selection of the penalty parameter on the LASSO MSE. The main results are not altered.

Note that in case $Z$ contains variables not included among the confounders $X$, there is some concern that including $Z$ in the estimation of the propensity score might inflate the propensity score without removing additional selection bias. Therefore, our main specification is based on $p(x)$ only. We also estimate specifications allowing $Z$ to enter the propensity score as well, $p(x, z)$ (see Appendix F.5). However, besides decreasing the precision of the estimates, the main results are not altered.

## 4.5 Estimation of CATEs

To avoid the situation in which the LASSO approach models idiosyncratic within-sample effects, we randomly partition the sample into two equal sized parts. We assume independence between the two samples. We use the first sample to select the relevant effect heterogeneity variables (training sample). We use the second sample for the estimation of a WOLS model including all selected heterogeneity variables (estimation sample). This is called the 'honest' inference procedure (see the discussion about the general properties, e.g., in Fithian, Sun, and Taylor, 2017).

The CATE for individual $i$ is estimated as $\hat{\gamma}(Z_i) = Z_i \hat{\delta}$. All coefficients of variables not selected in the training sample are set to zero. The coefficients of the selected variables are estimated in the estimation sample and extrapolated to the full sample. The medical and biometric literature calls $\hat{\gamma}(Z_i)$ individualised treatment effects (ITE) (e.g., Chen et al., 2017). The estimates of $\hat{\delta}$ vary with respect to the random sample split. To reduce the dependency of the results on a particular split, we run the analyses $S = 30$ times with different random splits. We calculate the individualised CATEs, $\hat{\gamma}_s(Z_i) = Z_i \hat{\delta}_s$, for each split, where the Post-LASSO coefficients, $\hat{\delta}_s$, are from the random sample split $s$. We use these parameters to calculate the aggregated CATEs, $\bar{\gamma}(Z_i) = \frac{1}{S}\sum_{s=1}^{S} \hat{\gamma}_s(Z_i)$. This procedure is in the spirit of bootstrap aggregation (`bagging`) in the machine learning literature (see, e.g., Breiman, 1996). It reduces model dependency and smooths the estimated CATEs, but the estimation model of $\bar{\gamma}(Z_i)$ is more difficult to interpret than the model of $\hat{\gamma}(Z_i)$. To understand which factors influence the aggregated CATEs, we report averages by different groups,

$$\bar{\gamma}_g = \frac{1}{\sum_{i=1}^{N} G_i} \sum_{i=1}^{N} G_i\, \bar{\gamma}(Z_i),$$

where the binary variable $G_i$ indicates whether individual $i$ belongs to the group ($G_i = 1$) or not ($G_i = 0$). These groups could, for example, be all JSP participants, all non-participants, or unemployed persons with specific characteristics.

## 4.6 Variance estimation

It appears natural to estimate the variance with a bootstrap approach over the whole estimation algorithm, including the variable selection step. However, this is computationally infeasible for a reasonable number of bootstrap replications. Thus, we use a computationally feasible bootstrap approach in which we fix the selected heterogeneity variables in each sample split.

First, we draw a random bootstrap sample $b$ (with replacement) clustered on the caseworker level. Second, for each sample split, we align the observations in the bootstrap sample to the observations in the original estimation sample. We only keep observations that we observe in both the bootstrap and the estimation sample. Third, based on these samples, we re-estimate the CATEs for each sample split using the heterogeneity variables selected in the original training sample of the respective sample split. We repeat these three steps 1,000 times. This procedure takes into account the dependencies that stem from overlapping observations across sample splits.

*Procedure 1: Estimation algorithm of the adapted MCM.*

| Step 1 | Estimate propensity score $\hat{p}(X_i, Z_i)$ and calculate the IPW weights. |
|---|---|
| Step 2 | a) Randomly split the sample into training and estimation sample $s$. <br> b) Select the relevant heterogeneity variables in the training sample using the LASSO approach with or without efficiency augmentation (explained in Appendix F.2). <br> c) Estimate the coefficients $\hat{\delta}_s$: <br>   (i) Set the coefficients of deselected variables to zero. <br>   (ii) Estimate the coefficients of the selected variables in the estimation sample. <br> d) Calculate $\hat{\gamma}_s(Z_i) = Z_i \hat{\delta}_s$ for the full sample. |
| Step 3 | a) Repeat Step 2 $S$ times. <br> b) Calculate the aggregated CATEs $\bar{\gamma}(Z_i) = \frac{1}{S}\sum_{s=1}^{S} \hat{\gamma}_s(Z_i)$ and group averages of CATEs $\bar{\gamma}_g = \frac{1}{\sum_{i=1}^{N} G_i} \sum_{i=1}^{N} G_i \bar{\gamma}(Z_i)$. |
| Step 4 | Bootstrap the variance of $\bar{\gamma}(Z_i)$ and $\bar{\gamma}_g$. (For computational feasibility, we do not re-estimate Step 2b) in the bootstrap replications.) |

For each sample split $s$ and bootstrap replication $b$ we obtain the bootstrapped CATEs, $\hat{\gamma}_{sb}(Z_i) = Z_i \hat{\delta}_{sb}$. The aggregated bootstrapped CATEs are $\bar{\gamma}_b(Z_i) = \frac{1}{S}\sum_{s=1}^{S} \hat{\gamma}_{sb}(Z_i)$. We estimate the standard error for the aggregated CATEs with

$$\hat{\sigma}_i = \sqrt{\frac{1}{B}\sum_{b=1}^{B}\left(\bar{\gamma}_b(Z_i) - \frac{1}{B}\sum_{b=1}^{B}\bar{\gamma}_b(Z_i)\right)^2},$$

and the standard errors of CATEs by groups with

$$\hat{\sigma}_g = \sqrt{\frac{1}{B}\sum_{b=1}^{B}\left(\bar{\gamma}_{gb} - \frac{1}{B}\sum_{b=1}^{B}\bar{\gamma}_{gb}\right)^2},$$

where $\bar{\gamma}_{gb}$ is the estimate of $\bar{\gamma}_g$ in the respective bootstrap replication $b$.

# 5 Results

## 5.1 Propensity score model

Table D.1 in Appendix D reports the average marginal effects of the estimated propensity score model. The propensity score estimates serve as inputs into the matching algorithm. The results

confirm the impression from the descriptive statistics in Table 1, namely that the participation probability is generally increasing with previous labour market success. Unemployed persons with good labour market opportunities have a greater probability to participate in a JSP. Such a selection of training participants is called 'cream-skimming' (e.g., Bell and Orr, 2002). The effect of training is not necessarily higher for participants with good labour market opportunities, because these participants would have good labour market opportunities even in the absence of training (see, e.g., the discussion in Berger, Black, and Smith, 2000).

When performing matching, it is a best practice to check for potential issues of (i) insufficient support in the propensity scores across treatment states that may result in incomparable matches as well as large matching weights of some non-treated observations with specific propensity scores; and (ii) imbalances in covariates after matching (due to inappropriate propensity score specifications). We document the distribution of the baseline propensity score in Figure D.1 of Online Appendix D. Furthermore, we document the balancing of the control variables after matching in Table D.2 of Online Appendix D. We find only small imbalances between JSP participants and non-participants. The standardised differences are always below three.

## 5.2 Average effects

Figure 1 shows the estimated potential outcomes and average programme effects on employment for each of the first 31 months after the programme's start. We observe substantial negative lock-in effects. The employment probability in the first three months is about 15 percentage points lower for JSP participants compared to non-participants. However, differences in the two groups' employment probabilities disappear after 16 months. In months 22 to 24 after a programme's start, we find small positive effects. But this seems to be only of short duration. Overall, the long-term effects are insignificant and close to zero. The negative lock-in effects are consistent with the findings of the previous Swiss JSP evaluations (e.g.,

Gerfin and Lechner, 2002, Lalive, van Ours, and Zweimüller, 2008). Moreover, the effectiveness of JSPs is also negative in other countries (see e.g., Dolton and O'Neil, 2002, Wunsch and Lechner, 2008). It is possible that participants reduce the intensity of informal job search during participation in a JSP, which could explain negative employment effects.

Searching for effect heterogeneity in each month after a programme's start is computationally expensive and hard to intuitively summarise (at least if it varies over time). Therefore, we estimate the effects of JSP participation on cumulated months employed during the first 6, 12, and 31 months after a programme begins, as well as during months 25 to 31. Table 2 shows the respective average effects that mirror the findings in Figure 1. The lower employment probabilities after programme participation translate into an average decline of 0.8 employment months ($\approx$ -24 days) during the first six months after the start of participation. This decreases to -1.1 months ($\approx$ -33 days) during the first 12 and 31 months. We find no significant employment effects during months 25 to 31 after the start of participation.

*Figure 1: ATE, ATET, and potential outcome levels by months since the start of JSP participation.*

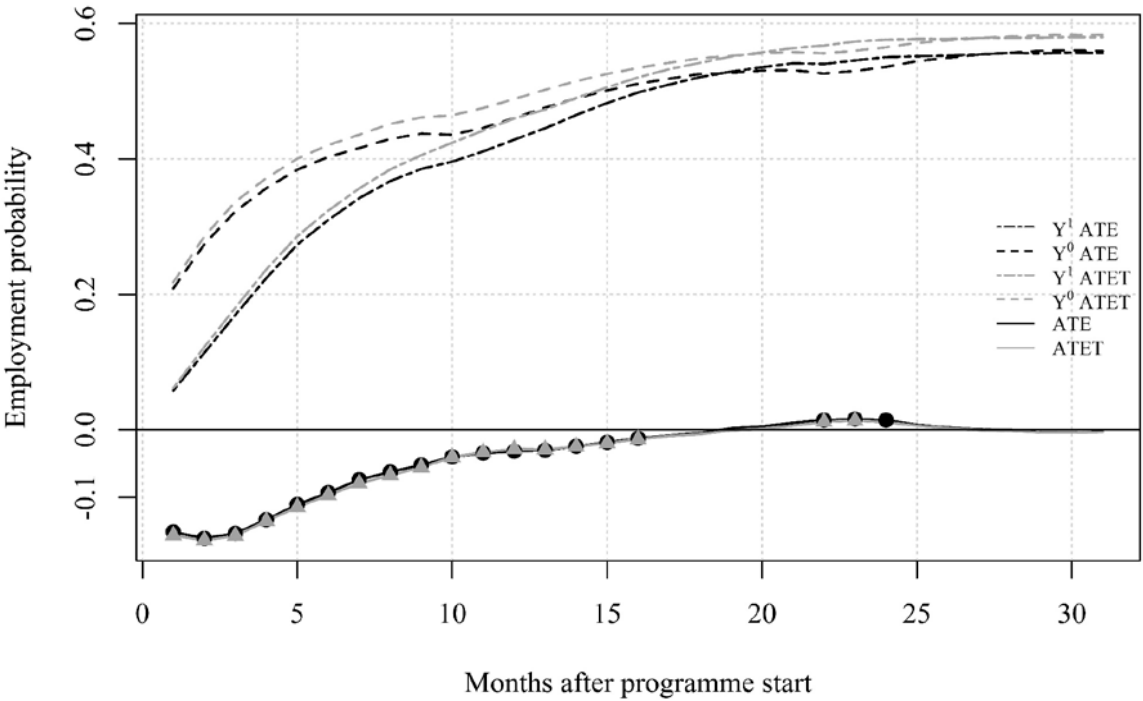

Note: We estimate the ATE and ATET separately for each of the first 31 months after the start of JSP participation. Circles/triangles indicate significant effects at the 5% level. We obtain standard errors from a clustered bootstrap at the caseworker level with 4,999 replications.

*Table 2: ATE, ATET, and ATENT by duration since the start of JSP participation.*

| Months employed since start of participation | ATE | | ATET | | ATENT | |
|---|---|---|---|---|---|---|
| | Coef. | S.E. | Coef. | S.E. | Coef. | S.E. |
| | (1) | | (2) | | (3) | |
| During first 6 months | -0.80*** | (0.02) | -0.82*** | (0.02) | -0.80*** | (0.02) |
| During first 12 months | -1.10*** | (0.05) | -1.13*** | (0.04) | -1.09*** | (0.05) |
| During first 31 months | -1.14*** | (0.14) | -1.20*** | (0.13) | -1.12*** | (0.15) |
| During months 25-31 | -0.007 | (0.03) | -0.011 | (0.03) | -0.007 | (0.04) |

Note: We obtain standard errors (S.E.) from a clustered bootstrap at the caseworker level with 4,999 replications. *, **, *** mean statistically different from zero at the 10%, 5%, 1% level, respectively.

## 5.3 Effect heterogeneity

Table 3 reports the estimated heterogeneity coefficients, $\hat{\delta}$, obtained from one of the considered random partitions into training and estimation samples.[15] The coefficients are the marginal effects of the respective variables on the treatment effect of JSP (as opposed to the marginal effects of the respective variables on the outcome level in standard linear regression models).

The first column of Table 3 reports the estimated coefficients for the outcome cumulated employment during the first six months after training participation begins. In this specification, the Post-LASSO estimation selects 17 out of 1,268 potential variables. In the estimation sample, five of these variables are significant; for example, the treatment effect increments by 0.3 months ($\approx$ 9 days) for unskilled workers with previous earnings below 25,000 CHF a year (see row 3). When all other selected variables equal zero, the predicted effect of JSP participation for unskilled workers with previous earnings below 25,000 CHF a year would be $-0.89 + 0.3 = -0.59$ months employment ($\approx$ -17 days). However, we must be cautious when interpreting the coefficients, because they are selected to maximise prediction power and might differ from the structural (causal) model (see, e.g., discussion in Mullainathan and Spiess, 2017).

---

[15] We omit the coefficients of the main effects because they are only used for efficiency augmentation and irrelevant for the interpretation.

*Table 3: Post-LASSO coefficients for selected outcome variables.*

|  | Months employed during first 6 months after the start of participation | | Months employed during first 12 months after the start of participation | |
|---|---|---|---|---|
|  | Coef. | S.E. | Coef. | S.E. |
|  | (1) | | (2) | |
| Constant | -0.89*** | (0.05) | -1.29*** | (0.09) |
| # of unemp. spells in last two years | 0.06 | (0.12) | - | - |
| Unskilled × past income 0 - 25k | 0.30*** | (0.11) | 0.53 | (0.53) |
| Skilled w/o degree × same gender like CW | 0.20 | (0.21) | - | - |
| Skilled w/o degree × age difference between unemployed & CW | -0.01 | (0.01) | - | - |
| # of unemp. spells in last 2 years × age of CW | 0.00 | (0.00) | - | - |
| # of unemp. spells in last 2 years × medium city size | -0.05 | (0.06) | -0.13 | (0.14) |
| # of unemp. spells in last 2 years × past income 0 - 25k | -0.04 | (0.06) | -0.10 | (0.14) |
| # of unemp. spells in last 2 years × prev. job unskilled | 0.04 | (0.05) | 0.21* | (0.13) |
| # of unemp. spells in last 2 years × same gender like CW | -0.01 | (0.05) | - | - |
| CW has own unemp. experience × prev. job unskilled | 0.19** | (0.09) | 0.34* | (0.21) |
| Foreigner with perm. residence permit × past income 25 - 50k | 0.19 | (0.12) | - | - |
| Small city × past income 50 - 75k | -0.16* | (0.09) | -0.26 | (0.20) |
| Single household × no emp. spell last 2 years | -0.17** | (0.08) | - | - |
| Single household × prev. job unskilled | 0.16 | (0.11) | - | - |
| Prev. job primary sector × age difference between unemp. person & CW | -0.02** | (0.01) | - | - |
| Prev. job restaurant | -0.01 | (0.12) | - | - |
| Prev. job tourist sector | -0.09 | (0.12) | - | - |
| Unskilled × prev. job unskilled | - | - | -0.22 | (0.64) |
| # of unemp. spells in last 2 years × unempl. & CW have primary education | - | - | 0.19** | (0.08) |
| CW has vocational training degree × past income 50 - 75k | - | - | -0.13 | (0.30) |
| Past income 25 - 50k × unskilled | - | - | 0.14 | (0.24) |
| # of emp. spells past 5 years × prev. job in primary sector | - | - | -0.24 | (2.16) |
| Prev. job in primary sector × unskilled | - | - | -0.19 | (0.53) |
| Regional emp. agency No. 44 | - | - | -0.68 | (0.52) |
| # of selected variables | 17 of 1,268 | | 13 of 1,268 | |

Note: We apply one-step efficiency augmentation. We partition the data randomly into selection and estimation samples. We choose the penalty term based on Post-LASSO RMSE, which we optimise with 10-fold cross-validation. We obtain standard errors (S.E.) from a clustered bootstrap at the caseworker level with 4,999 replications. *, **, *** mean statistically different from zero at the 10%, 5%, 1% level, respectively. We report results for additional outcomes in Table E.1 of Online Appendix E. CW is the abbreviation for caseworker. 25 - 50k is the abbreviation for 25,000-50,000 CHF. 50 – 75k is the abbreviation for 50,000-75,000 CHF.

The second column of Table 3 shows the coefficients for the thirteen selected heterogeneity variables for the outcome cumulated employment during the first twelve months after training participation begins. The selected heterogeneity variables are partially overlapping between the two outcomes (comp. column 1 and 2). In Table E.1 in Online Appendix E, we report the selected heterogeneity parameters for the outcome cumulated months employed in the first 31 months after training participation begins. We omit the results for the outcome cumulated months employed between months 25 to 31 after training participation begins, because we do not detect any effect heterogeneity in the considered sample split.

To improve precision and check the sensitivity of our results, we investigate the Post-LASSO models for different random sample splits. For each random partition, we obtain different Post-LASSO models (Table F.6 in Online Appendix F documents the number of selected variables in the different random sample splits). This is unsurprising, because many of the variables we consider are highly correlated (e.g., different measures of the employment history). Therefore, the same CATE can be obtained from different Post-LASSO models, each considering different variables or different functions of variables. Table F.1 in Online Appendix F documents the average correlation between CATEs for different sample splits. The correlations are positive and relatively large. Accordingly, the CATEs are consistent across the sample splits we consider. The selected models are not identical, but each model essentially predicts similar CATEs.

One approach to get an overview of the heterogeneities we detect is to plot the distribution of the predicted effects. Therefore, Figure 2 reports the distribution of the aggregated CATEs of JSPs on cumulated months employed during the first six months after participation begins. The figure documents substantial variation in the aggregated CATEs. For most groups of unemployed persons the aggregated CATE of JSP participation is between -0.8 and -1 months of employment (approximately a decline of between 24 and 30 days). However, the CATEs are less negative or even positive for a non-negligible fraction of the unemployed persons. This points at potential ways to improve assignment to a JSP.

*Figure 2:  Distribution of aggregated CATEs for months employed during first six months after the start of participation.*

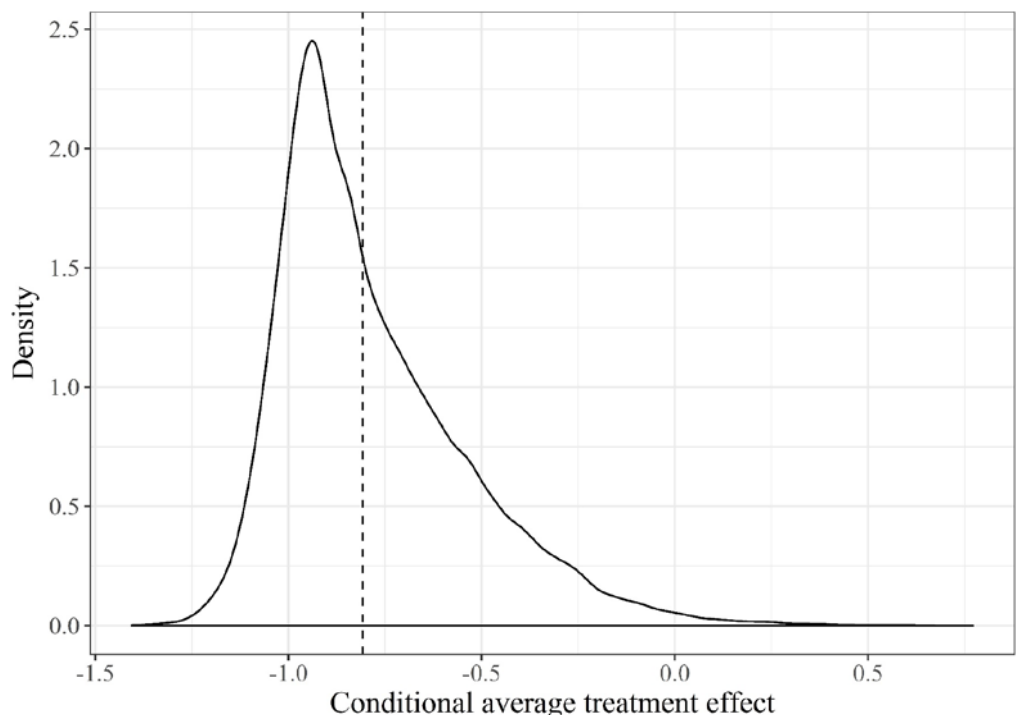

Note:   Kernel smoothed distribution of CATEs. Gaussian kernel with bandwidth 0.02, chosen by Silverman's rule-of-thumb. We apply one-step efficiency augmentation. We partition the data randomly into selection and estimation samples. We choose the penalty term based on Post-LASSO RMSE, which we optimise with 10-fold cross-validation. The dashed vertical line shows the ATE.

Table 4 reports summary statistics for the aggregated CATEs. For all outcomes, the means of the aggregated CATEs are close to the (semi-parametrically) estimated ATEs (comp. Table 2). This confirms that the estimation of the aggregated CATEs works well, on average. For all outcomes, the median is slightly lower than the mean. This suggests a right-skewed distribution (similar to Figure 2). We find substantial heterogeneity for the outcomes cumulated months employed during the first 12 months and the first 31 months after the start of JSP participation. After 12 months, the JSP effect ranges from minus two to plus two employment months. After 31 months, the JSP effect ranges from minus three to plus three employment months. However, for the outcome cumulated months employed between month 25 and 31 after the start of JSP participation we find little heterogeneity.

*Table 4: Descriptive statistics of aggregated CATEs.*

| Months employed since start of participation | Mean | Median | S.D. | Min. | Max. | Mean S.E. |
|---|---|---|---|---|---|---|
| | (1) | (2) | (3) | (4) | (5) | (6) |
| During first 6 months | -0.78 | -0.84 | 0.25 | -1.41 | 0.77 | 0.07 |
| During first 12 months | -1.10 | -1.20 | 0.32 | -2.09 | 1.44 | 0.10 |
| During first 31 months | -1.13 | -1.25 | 0.60 | -3.79 | 4.12 | 0.23 |
| During months 25-31 | -0.04 | -0.05 | 0.06 | -0.32 | 0.48 | 0.04 |

Note: We obtain CATEs from aggregating CATEs from 30 different random sample splits. Standard deviations are abbreviated with S.D. in column (3). Column (6) shows mean standard errors of CATEs.

Accordingly, the MCM successfully discovers substantial effect heterogeneity. However, interpretation of the results is not straightforward, because the underlying functions are too complex. Figure 2 and Table 4 document two ways to aggregate the results. However, we want to go beyond these abstract descriptions and make explicit policy recommendations. In the next section, we marginalise the effects for specific variables of interest. This enables us to reveal more of the CATEs' structure. Afterwards, we focus on the implementation of specific JSP assignment rules.

## 5.4 Effect heterogeneity by selected variables

In this section, we average CATEs by characteristics of unemployed persons and their case-workers. For each characteristic, we partition the sample into two mutually exclusive groups (high $g = 1$ and low $g = 0$ group), by using a binary characteristic itself or by discretising at the median of non-binary characteristics. The parameters $\bar{\gamma}_{g=1}$ and $\bar{\gamma}_{g=0}$ average the CATEs over all unemployed in the respective group.

Figure 3 reports effect heterogeneity of JSP participation on cumulated months employed during the first six months after the start of participation by low and high values of the characteristics of unemployed persons. The groups in the top of Figure 3 show the largest effect heterogeneities. For groups at the bottom of Figure 3 we find only little effect heterogeneity. We estimate the largest degree of effect heterogeneity for unskilled workers. The average effect of JSP for the unskilled unemployed is 0.26 months (≈ 8 days) longer than for unemployed persons in other skill categories (see Table E.2 in Online Appendix E).

*Figure 3: CATEs on cumulated employment during the first 6 months after start JSP participation by characteristics of unemployed persons.*

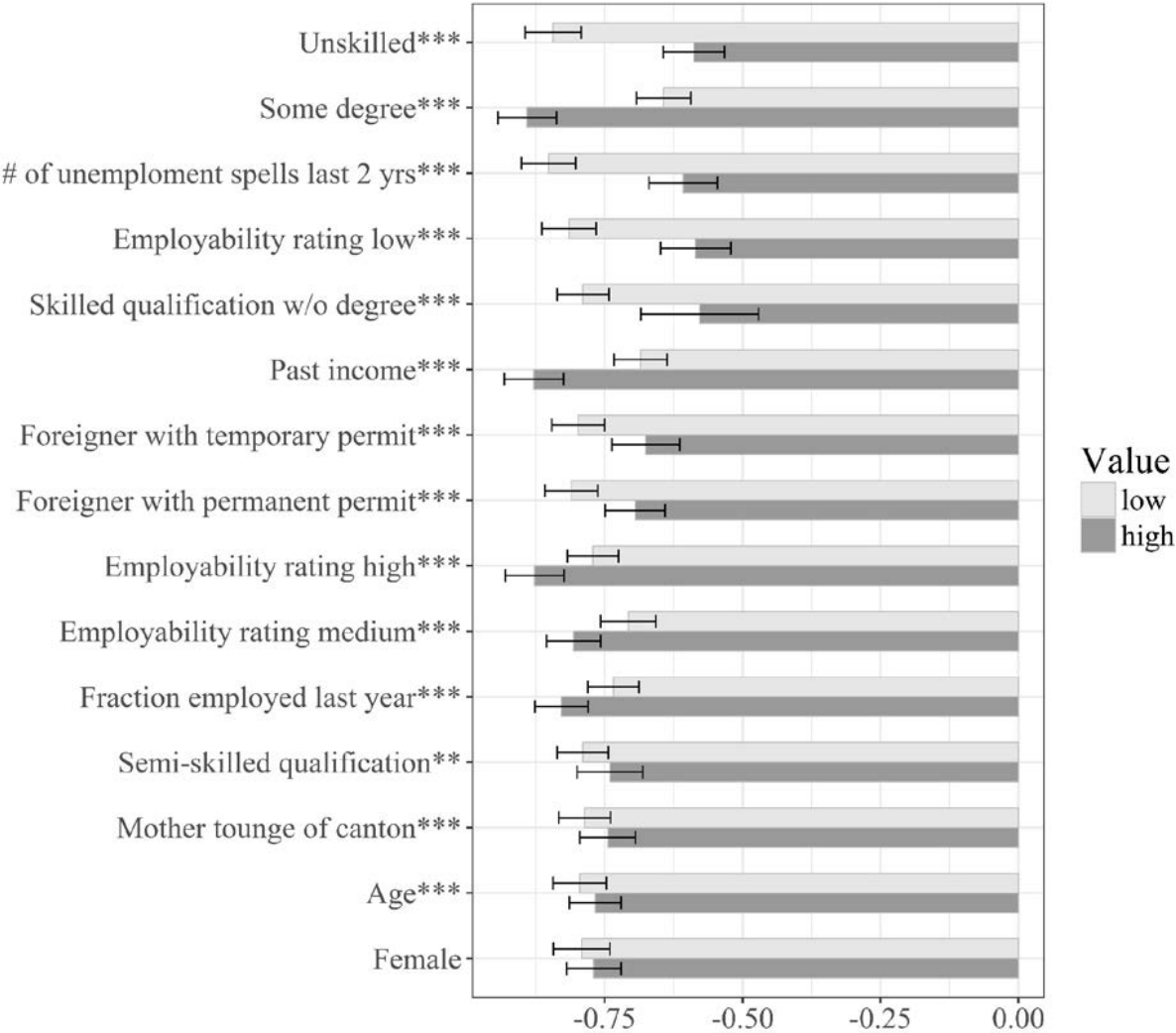

Note: CATEs by low and high values of the respective characteristic of unemployed persons. A low value is zero when the variable is binary or below the median when the variable is non-binary. A high value is one when the variable is binary or not below the median when the variable is non-binary. The CATEs are based on 30 random sample splits. For each partition, we choose the penalty term based on Post-LASSO RMSE, which we optimise with 10-fold cross-validation. We apply one-step efficiency augmentation. We report the 95%-confidence interval based on the bootstrap procedure described in section 4.6. *, **, *** mean statistically different from zero at the 10%, 5%, 1% level, respectively. The differences and respective standard errors are reported in Table E.2 in Online Appendix E. We report results for additional outcomes in Figures E.1, E.3, and E.5 in Online Appendix E.

Conversely, Figure 3 documents that individuals with some degree of education suffer on average more from JSP than individuals with no degree. In general, we observe that the negative lock-in effect is much less pronounced for unemployed persons with lesser qualifications. This suggests that cream-skimming reduces the effectiveness of JSP participation. These findings are consistent with the evaluation literature (e.g., Card, Kluve, and Weber, 2015, van den Berg and van der Klaauw, 2006). Furthermore, the lock-in effects are less negative for foreigners. One potential explanation is that foreigners have a relatively small network for an informal job search. Therefore, the formal job search strategy might be relatively successful for them. This suggests more foreigners should be assigned to JSPs. We find only a little heterogeneity by gender and age, which is in line with the findings of Vikström, Rosholm, and Svarer (2013) for JSPs in Denmark. In our application, the effect heterogeneity by gender is not statistically significant (see standard errors in Table E.2 in Online Appendix E).

Figure 4 reports effect heterogeneity of JSP participation on cumulated months employed during the first six months after the start of participation by low and high values of caseworker characteristics. The interpretation of Figure 4 corresponds to the interpretation of Figure 3. Although we find some statistically significant differences, they are much less pronounced than for the characteristics of unemployed persons. Most effect heterogeneity is observed by caseworkers' own unemployment experience, but the difference is only 0.07 months ($\approx$ 2 days). However, the difference is statistically significant (see Table E.3 in Online Appendix E). Interestingly, the cooperativeness of caseworkers has no statistically significant influence on the effectiveness of JSP participation. We would have expected this characteristic to be a good predictor for effect heterogeneity, because it might approximate different monitoring intensities of the caseworker.

*Figure 4: CATEs on cumulated employment during the first 6 months after start JSP participation by caseworker characteristics.*

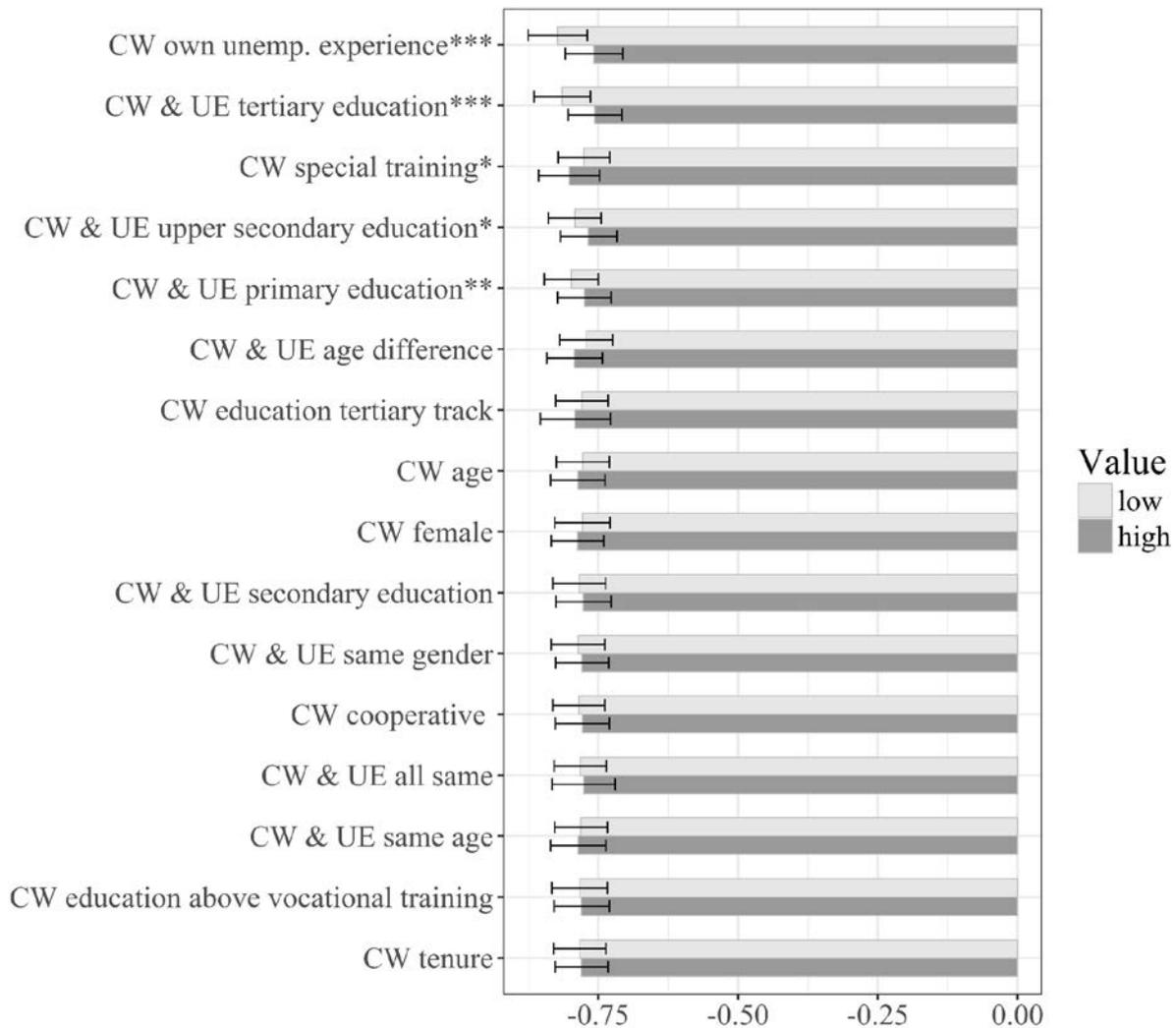

Note: CATEs by low and high values of the respective caseworker characteristic. A low value is zero when the variable is binary or below the median when the variable is non-binary. A high value is one when the variable is binary or not below the median when the variable is non-binary. The CATEs are based on 30 random sample splits. For each partition, we choose the penalty term based on Post-LASSO RMSE, which we optimise with 10-fold cross-validation.. We apply one-step efficiency augmentation. We report the 95%-confidence interval based on the bootstrap procedure described in section 4.6. *, **, *** mean statistically different from zero at the 10%, 5%, 1% level, respectively. The differences and respective standard errors are reported in Table E.3 in Online Appendix E. We report results for additional outcomes in Figures E.2, E.4, and E.6 in Online Appendix E. CW is the abbreviation for caseworker.

## 5.5 Assignment rules for JSP

Next, we investigate the characteristics of unemployed persons with positive CATEs (Table 5). The number of individuals with positive CATEs amounts to 674, which corresponds to 0.9% of the unemployed persons in the sample. The first row of Table 5 reports the share of unemployed persons assigned to a JSP by the sign of the CATEs. Only 7% of unemployed persons with positive CATEs participate in a JSP, whereas 16% of the unemployed persons with negative CATEs participate in a JSP. This points to the potential to improve the selection of JSP participants. Figure 5 provides further evidence in this direction. It documents the relationship between the CATEs and the propensity score. An effective selection of JSPs would show a positive relationship, indicating that those who benefit more have a higher JSP participation probability. However, Figure 5 shows over a wide range (i.e., for unemployed persons with more than 10% participation probability) a negative relationship between the CATEs and the propensity score. This implies, those unemployed persons who suffer more from JSPs have a higher participation probability.

*Figure 5: Correlation of the propensity score and the CATEs.*

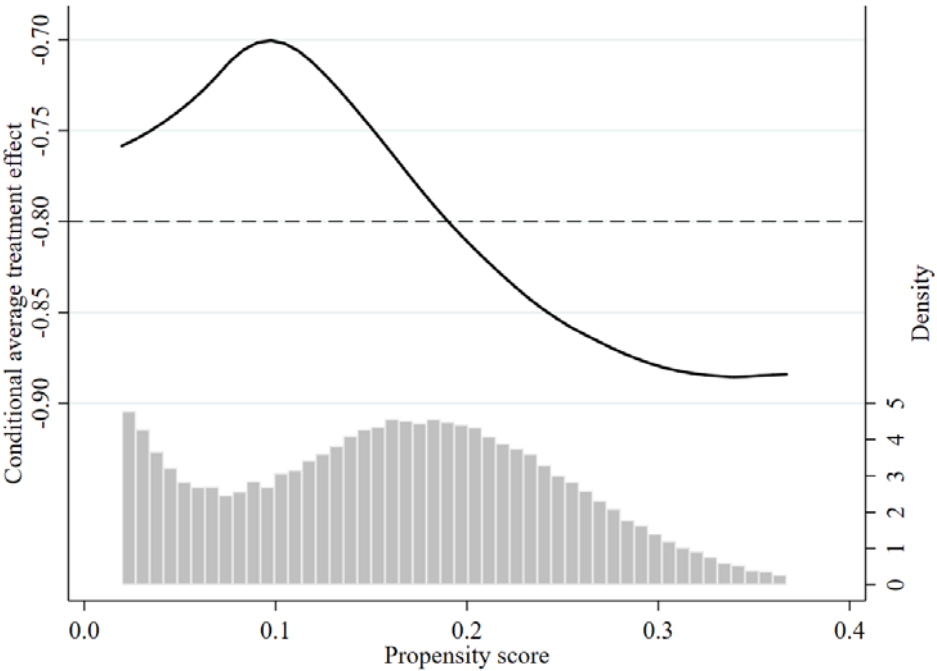

Note: Kernel smoothed regression of propensity score and CATEs. Local constant kernel regression used with Gaussian kernel and bandwidth 0.02 computed by the plug-in bandwidth selector Fan and Gijbles (1996). The dashed horizontal line shows the ATE. The grey bars show the histogram of the propensity score.

Additionally, Table 5 reports characteristics of unemployed persons with positive and negative CATEs. The difference gives explicit advice on how assignment rules to JSPs could be improved. For example, unemployed persons with a lower past income and lesser past employment experience tend to have positive effects from participation. Participants with lower degrees of education and foreigners seem to have a higher probability to profit from a JSP. Strikingly, those unemployed persons who receive a low employability rating by their caseworker are more likely to experience positive effects from a JSP than unemployed persons with a medium or high rating. These results are further evidence that cream-skimming does not improve JSP effectiveness.

Furthermore, we document the effectiveness of hypothetical statistical assignment rules in Table 6. Statistical assignment rules have already received considerable attention in the context of ALMPs (see, e.g., Bell and Orr, 2002, Caliendo, Hujer, and Thomsen, 2008, Frölich, 2008, Dehejia, 2005, O'Leary, Decker, and Wandner, 2002, among many others). However, we are not aware of any application using machine learning methods to investigate assignment rules for ALMP that systematically consider a high-dimensional covariate space.

For the proposed assignment rules, we keep the number of 12,712 (hypothetical) JSP participants constant.[16] Therefore, the proposed assignment rules are (almost) cost-neutral compared to the existing assignment mechanism. However, we do not account for possible capacity limits in regional training centres. We consider five hypothetical assignment rules: (i) random allocation (called 'random' in the following), (ii) assignment of unemployed persons with the highest CATEs (called 'best case' in the following), (iii) assignment of unemployed persons with the lowest CATEs (called 'worst case' in the following), (iv) all unemployed persons with at least one unemployment spell in the previous two years and unskilled plus a random selection of the remaining unemployed persons with at least one unemployment spell in the previous two years and no degree (called 'previous unemployment' in the following), and (v) all unemployed with a low employability rating by their caseworkers plus a random sample with a medium employability rating (called 'employability rating' in the following). The random adding of participants in assignment rules (iv) and (v) enables us to maintain the number of 12,712 participants. The 'previous unemployment' rule (iv) is inspired by the variables that show the highest treatment effects in Table 3 and Figure 3. The 'employability rating' rule (v) assigns unemployed persons to a JSP for whom the caseworkers give a low employability rating, as opposed to cream-skimming, which assigns more unemployed persons with high employability ratings.

---

[16] We consider only participants on the common support. Therefore, the number of participants considered here is lower than previous numbers.

Table 6 reports the average CATE under the different assignment rules. The average CATE represents the hypothetical ATET under this treatment assignment. The 'worst case' and 'best case' assignment rules are the lower and upper bounds of the ATET (for a fixed number of 12,712 participants). The difference between the lower and upper bounds are about 0.65 employment months ($\approx$ 20 days). The ATET under random assignment is -0.78 months ($\approx$ -24 days) employment during the first six months after the start of participation. This is the benchmark assignment rule. Any imposed assignment rule should be better than random assignment. However, the observed ATET is -0.82 months ($\approx$ -25 days) employment during the first six months after the start of the programme. It appears that the current assignment mechanism is not better than a random assignment rule. In the context of Swiss ALMPs, Lechner and Smith (2007) also find that the allocation by caseworkers performs no better than random assignment. Furthermore, this is consistent with the findings of Bell and Orr (2002) and Frölich (2008), who reject the idea that caseworkers allocate training programs efficiently in the US and Sweden. Applying the optimal assignment rule 'best case' would reduce the negative employment effects by 60% $(= \big((0.82 - 0.33)/0.82\big) \cdot 100\%)$.

For the proposed assignment rule 'previous employment' the predicted ATET is -0.51 months ($\approx$ -15 days) employment during the first six months after the start of participation. On average, each participant has 9 days more employment under this assignment rule than under random assignment. The negative employment effect of the current assignment mechanism would be reduced by 38% $(= ((0.82 − 0.51)/0.82) \cdot 100\%)$. For the proposed assignment rule 'employability rating' the predicted ATET is -0.61 months ($\approx$ -18 days) employment during the first six months after the start of participation. On average, each participant has 6 days more employment under this assignment rule than under random assignment. The negative employment effect of the current assignment mechanism would be reduced by 21% $(= ((0.82 − 0.61)/0.82) \cdot 100\%)$. These results are consistent with the argument that assignments based on expected treatment effects rather than on predicted outcomes can be more successful (Ascarza, 2016). However, the average effects remain negative and the programme does not seem useful in improving the employment opportunities of unemployed persons in general. Nevertheless, the easy-to-implement assignment rules document the potential to improve the current allocation mechanism.

*Table 5: Characteristics of unemployed by the sign of CATE.*

|  | $\bar{\gamma}_i \geq 0$ | $\bar{\gamma}_i < 0$ | Difference | S.E. |
|---|---|---|---|---|
|  | (1) | (2) | (3) | (4) |
| JSP participation | 0.07 | 0.16 | -0.09*** | (0.01) |
| Female | 0.41 | 0.45 | -0.04 | (0.07) |
| Past income (in 10,000 CHF) | 0.32 | 0.42 | -0.11*** | (0.02) |
| Fraction of months emp. in last 2 years | 0.70 | 0.80 | -0.10*** | (0.01) |
| # of unemp. spells in last 2 years | 4.71 | 0.54 | 4.17*** | (0.50) |
| Unskilled | 0.62 | 0.24 | 0.38*** | (0.09) |
| Semiskilled | 0.16 | 0.16 | 0.00 | (0.05) |
| Skilled without degree | 0.14 | 0.04 | 0.10* | (0.06) |
| Some educational degree | 0.08 | 0.57 | -0.49*** | (0.03) |
| Foreigner with mother tongue is cantons' language | 0.14 | 0.11 | 0.03 | (0.02) |
| Low employability rating by CW | 0.43 | 0.14 | 0.29*** | (0.11) |
| Medium employability rating by CW | 0.56 | 0.76 | -0.20* | (0.11) |
| High employability rating by CW | 0.01 | 0.10 | -0.10*** | (0.003) |
| Age (in 10 years) | 3.57 | 3.67 | -0.10 | (0.08) |
| Foreigner with temporary residence permit | 0.32 | 0.13 | 0.19*** | (0.06) |
| Foreigner with permanent residence permit | 0.41 | 0.25 | 0.16** | (0.07) |
| # of individuals | 674 | 77,824 |  |  |

Note: Average characteristics of individuals with positive and negative CATE in the first 6 months after start of participation. The CATEs are based on 30 random sample splits. For each partition, we choose the penalty term based on Post-LASSO RMSE, which we optimise with 10-fold cross-validation. We apply one-step efficiency augmentation. We report the 95%-confidence interval based on the bootstrap procedure described in section 4.6. *, **, *** mean statistically different from zero at the 10%, 5%, 1% level, respectively. CW is the abbreviation for caseworker.

*Table 6: Average CATE of hypothetical participants under different assignment rules.*

| Assignment rule | CATE for participants |
|---|---|
| | (1) |
| Observed (ATET) | -0.82 |
| (i) Random | -0.78 |
| (ii) Best case | -0.33 |
| (iii) Worst case | -1.07 |
| (iv) Previous unemployment | -0.51 |
| (v) Employability rating | -0.61 |
| # of participants | 12,712 |

Note: Based on CATEs of 30 replications with one step efficiency augmented 10-fold cross-validated Post-LASSO.

## 5.6 Sensitivity checks

We perform large-scale sensitivity analyses to investigate the robustness of our results with respect to the choice of the empirical method and the selection of tuning parameters. We replicate our estimates using different forms of efficiency augmentation (see Online Appendix F.2). As an alternative variable selector, we consider the adaptive LASSO (Zou, 2006, see Online Appendix F.3). Furthermore, we replicate the results with the Modified Outcome Method (MOM) (Signorovitch, 2007, Zhang et al., 2012, see Online Appendix F.1) instead of the MCM. Moreover, we employ radius-matching with bias adjustment (Lechner, Miquel, and Wunsch, 2011) to balance the observable covariates between the treatment and control group instead of the IPW weights. This method shows good finite sample performance (Huber, Lechner, and Wunsch, 2013). Furthermore, we compare the robustness of the main results with two different sets of additional confounders (see Online Appendix F.5 for a description how we select the additional confounders). Finally, we compare our results with the causal forest approach (Wager and Athey, 2017, see Online Appendix F.4).

Table 7 reports the correlation between the CATEs for different empirical procedures. No matter which specification we use, the correlation between the CATEs is always positive and mostly above 0.5. The causal forest CATEs are less strongly correlated, but they still show a decently strong positive association. Accordingly, our main findings are not overly sensitive to the choice of empirical methods or selection of tuning parameters. We report additional sensitivity checks in Online Appendix F.6. The estimation results are widely consistent across a variety of different methodological choices and estimation procedures.

*Table 7: Correlation between CATEs obtained from different empirical procedures.*

| Cumulated employment during first 6 months | (1) | (2) | (3) | (4) | (5) | (6) |
|---|---|---|---|---|---|---|
| (1) MCM, one-step EA, Post-LASSO | 1.00 | | | | | |
| (2) MCM, two-step EA, Post-LASSO | 0.87 | 1.00 | | | | |
| (3) MCM, no EA, Post-LASSO | 0.77 | 0.77 | 1.00 | | | |
| (4) MCM, one-step EA, adaptive LASSO | 0.78 | 0.55 | 0.62 | 1.00 | | |
| (5) MCM, two-step EA, adaptive LASSO | 0.77 | 0.56 | 0.58 | 0.87 | 1.00 | |
| (6) MCM, no EA, adaptive LASSO | 0.67 | 0.56 | 0.83 | 0.67 | 0.62 | 1.00 |
| (7) MOM, Post-LASSO | 0.75 | 0.77 | 0.81 | 0.58 | 0.56 | 0.64 |
| (8) MOM, adaptive LASSO | 0.62 | 0.49 | 0.66 | 0.67 | 0.72 | 0.71 |
| (9) MCM, one-step EA, Post-LASSO with radius-matching weights | 0.97 | 0.87 | 0.76 | 0.74 | 0.73 | 0.66 |
| (10) MCM, one-step EA, LASSO | 0.85 | 0.61 | 0.65 | 0.93 | 0.79 | 0.62 |
| (11) Procedure (1) + additional confounders 1 | 0.83 | 0.75 | 0.65 | 0.59 | 0.62 | 0.54 |
| (12) Procedure (11) + additional confounders 2 | 0.90 | 0.86 | 0.72 | 0.62 | 0.65 | 0.59 |
| (13) Causal forest | 0.55 | 0.47 | 0.46 | 0.47 | 0.44 | 0.50 |
| Cumulated employment during first 6 months | (7) | (8) | (9) | (10) | (11) | (12) |
| (8) MOM, adaptive LASSO | 0.55 | 1.00 | | | | |
| (9) MCM, one-step EA, Post-LASSO with radius-matching weights | 0.74 | 0.61 | 1.00 | | | |
| (10) MCM, one-step EA, LASSO | 0.62 | 0.58 | 0.81 | 1.00 | | |
| (11) Procedure (1) + additional confounders 1 | 0.69 | 0.51 | 0.82 | 0.67 | 1.00 | |
| (12) Procedure (11) + additional confounders 2 | 0.77 | 0.55 | 0.91 | 0.70 | 0.92 | 1.00 |
| (13) Causal forest | 0.43 | 0.40 | 0.55 | 0.51 | 0.50 | 0.52 |

Note: Correlations of CATEs for different methods of efficiency augmentation, variable selection, modifications and weights. EA is the abbreviation for efficiency augmentation. If not specified differently, IPW weights are used to balance the covariates. In procedure (9), we use radius-matching weights (Lechner Miquel, and Wunsch, 2011). See Online Appendix F for more details about the different procedures. In Online Appendix F.5, we describe how we select additional confounders for procedures (11) and (12). Tables F.2-F.4 in Online Appendix F contain the correlation between CATEs for the other outcomes.

# 6 Conclusion

We investigate recently developed machine learning methods to uncover systematically treatment effect heterogeneity. We apply these methods to estimate the heterogeneous effects of Swiss Job Search Programmes (JSPs) on different employment outcomes by allowing for a high-dimensional set of variables potentially related to effect heterogeneity. We develop easy-to-implement, efficiency-improving assignment rules for JSPs.

The employment effects of JSPs are negative during the first six months after the start of participation and taper off afterwards. Parallel to this finding, we discover substantial effect heterogeneity during the first six months after the start of participation, but not afterwards. While an appropriate assignment rule could substantially decrease the negative lock-in effects, the negative effects are unlikely to disappear completely. In particular, we find that unemployed persons with low employment opportunities as well as foreigners experience less negative effects. The data used contains the caseworkers' subjective employability rating of their clients. Using this measure alone for programme assignment, i.e. if caseworkers assign mainly unemployed persons with a low employability rating, then negative lock-in effects are already reduced by approximately 22%. The results remain consistent across a range of alternative estimators and different implementation choices, showing the robustness of the findings.

There are still many open questions that are, however, beyond the scope of this paper. On the substantive side, for example, it is not clear that the largely negative results will generalize to other economic environments and other versions of JSPs implemented in other times and other countries. On the methodological side, it must be acknowledged that despite the extensive robustness checks, these methods are still very new and there could be practical problems not yet uncovered. We investigate the heterogeneous employment effects of a particular programme for different unemployed persons. The study abstracts from the questions about an optimal programme for a particular unemployed person, which is also relevant because of the usually rich programme structure of ALMPs. Such a modified goal raises several additional statistical issues that may be addressed in future research.